\documentclass[journal]{IEEEtran}

\usepackage{mathrsfs} 
\usepackage{xcolor}

\ifCLASSINFOpdf
   \usepackage[pdftex]{graphicx}
\else
   \usepackage[dvips]{graphicx}
\fi
\usepackage{amsmath,amssymb}
\usepackage{array}
\usepackage{fixltx2e}
\usepackage{stfloats}

\begin{document}

\makeatletter
\newcommand{\vast}{\bBigg@{4}}
\newcommand{\Vast}{\bBigg@{5}}
\makeatother
\title{On the assessment of an optimized method to determine the number of turns and the air gap length in ferrite-core low-frequency-current biased inductors}

\author{Andr\'es~Vazquez~Sieber
        and~M\'onica~Romero
\thanks{A. Vazquez Sieber and M. Romero are with Departamento de Electr\'onica, Facultad de Ciencias Exactas, Ingenier\'ia y Agrimensura, Universidad Nacional de Rosario, Rosario,
Santa Fe, 2000 Argentina e-mail: \{avazquez, mromero\}@fceia.unr.edu.ar.}}

\maketitle
\begin{abstract}
This paper presents a first assessment of a design method \cite{MyPaper:AVS_MR} aiming at the minimization of the number of turns $N$ and the air gap length $g$ in ferrite-core based low-frequency-current biased AC filter inductors. Several design cases are carried on a specific model of Power Module (PM) core, made of distinct ferrite materials and having different kinds of air gap arrangements The correspondingly obtained design results are firstly compared with the classic approach by linearization of the magnetic curve to calculate $N$ and the use of a fringing factor to determine $g$. Next, a refined design approach of specifying the inductance roll-off at the peak current and its potential limitations are discussed with respect to \cite{MyPaper:AVS_MR}. Finally, the behaviour of inductors designed according to \cite{MyPaper:AVS_MR} operated beyond their design specifications is analyzed.     
\end{abstract}

\IEEEpeerreviewmaketitle

\section{Introduction}

\IEEEPARstart{F}{errite} core based inductors are being increasingly employed into high-frequency high-power converters. Their low loss figures at such frequencies, their vast availability in shapes, sizes and materials already tailored for specific applications as well as their mature and well-known technology mainly justify that trend. To design a ferrite-based low-frequency-biased AC filter inductor is more challenging than a pure AC inductor since in the former case, the required inductance to block relatively high-frequency currents, has to be maintained even when usually a higher level of relatively low-frequency currents is superimposed. Complying the inductor with those constrains, \cite{MyPaper:AVS_MR} presents a design method that further determines the minimum number of turns to be winded and the optimum air gap length to be made on a given model of ungapped ferrite core. This is a very convenient strategy to minimize the copper power loss and simultaneously to reduce the intra-winding stray capacitance. It is therefore necessary a) to assess the results yield by \cite{MyPaper:AVS_MR} through comparisons with other design approaches and b) to establish its benefits and potential limitations.
\begin{figure}
				\includegraphics[width=8.5cm]{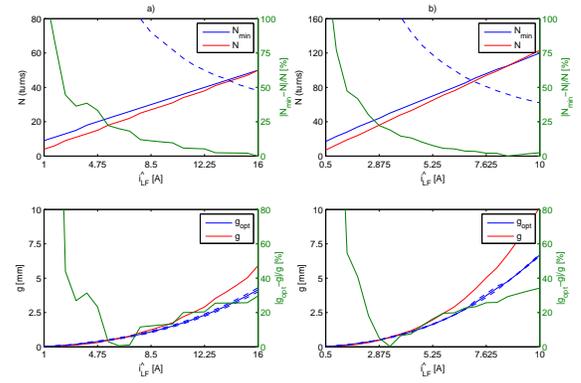} 
			\caption{$\{N_{min}, g_{opt}\}$ and $\{N, g\}$ as a function of $\hat i_{LF}$, for a constant $L_{\hat{rev}}$: a) 0.5mH, b) 2mH. N27 ferrite material, $q_g=1$.}
       \label{fig:Experiment_NvsiLF_L_1G_N27}
\end{figure}
\begin{figure}
				\includegraphics[width=8.5cm]{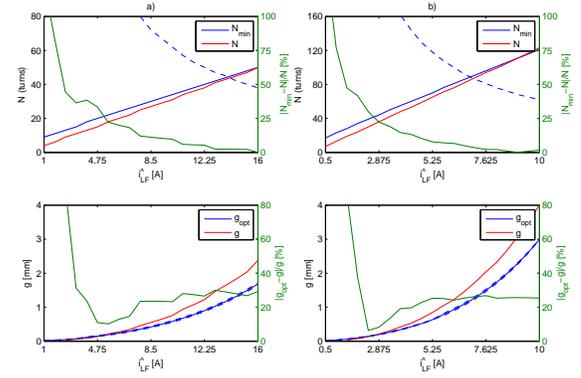} 
			\caption{$\{N_{min}, g_{opt}\}$ and $\{N, g\}$ as a function of $\hat i_{LF}$, for a constant $L_{\hat{rev}}$: a) 0.5mH, b) 2mH. N27 ferrite material, $q_g=2$.}
       \label{fig:Experiment_NvsiLF_L_N27}
\end{figure}

Although the design procedure of \cite{MyPaper:AVS_MR} can be applied to an inductor based on any kind of ferrite core, it is clearly oriented to high-power inductors. Any reduction in the number of winding turns translates into important savings in cost and thermal stress as well as the provision of a minimum, precise and accurate air gap length minimizes the heating and radiation effects of fringing fluxes while it facilitates the final inductance adjustment during manufacturing. Accordingly, the assessment of \cite{MyPaper:AVS_MR} should be firstly performed on inductors having core shapes well-suited to that power levels and so this paper is a first attempt in that direction. Among the many existing ferrite cores designed to handle relatively high currents, the Power Module (PM) core \cite{PMCores:EPCOS} is an attractive shape because a) it provides a balanced trade-off between magnetic shielding and power dissipation capabilities, b) it can be conveniently potted into a metallic heatsink to reduce the thermal resistance, c) its circular coil former offers a lower mean turn length than in a rectangular E core for the same cross-sectional area and d) it is easier to wind and to adjust the air gap compared to toroids. Hence, the assessment carried out in this paper focalizes on this type of core, but due to space restrictions, it is limited to a particular PM core model, the TDK-EPCOS PM 62/49 \cite{PMCores:EPCOS}. However, it is examined in combination with two different ferrite material, N27 and N87, as well as having two different air gap arrangements: a) a single air gap located in the central leg or b) three air gaps of equal length each one located in the respective leg of the core. The detailed definitions and datasheet parameters required to apply \cite{MyPaper:AVS_MR} to this particular model or to any other size/brand of PM ferrite cores are found in \cite{ReporteInternoPowerModuleCores:AVS}. Likewise, datasheet and model parameters related to ferrite material N27 and N87 needed by \cite{MyPaper:AVS_MR} are available in \cite{ReporteInternoDefinitionsBackground:AVS}.

This paper is organized as follows. In section \ref{sec:Fundamentals}, the fundamentals of \cite{MyPaper:AVS_MR} are revisited and the general setup of the simulations are presented. In section \ref{sec:Linear_Fringing}, the design results yields by \cite{MyPaper:AVS_MR} are compared with that obtained from the traditional method by linearization of the ferrite magnetization curve plus the use of a fringing factor to determine the air gap. In section \ref{sec:Roll-off}, a conceptual critique is posed on a refined method which requires to specify the initial inductance and its roll-off at peak current. In section \ref{sec:Beyond_specifications}, inductors designed according to \cite{MyPaper:AVS_MR} are driven beyond their original specifications and their behaviours are then analyzed. In section \ref{sec:Conclusion} conclusions are presented. For the sake of completeness, definitions of variables and parameters belonging to the models and methods of \cite{MyPaper:AVS_MR}, which are referred all along this paper are summarized in Appendix \ref{sec:Appendix}.

\section{Fundamentals and setup of the design method} \label{sec:Fundamentals}

In this section the key concepts and requirements of the method developed in \cite{MyPaper:AVS_MR} are refreshed as well as the general setup and some particular implementation details are explained.

The goal of the design method presented in \cite{MyPaper:AVS_MR} is the obtention of an inductor with the minimum number of turns $N_{min}$ and the minimum nominal air gap length with its tolerance $g=g_{opt}\pm \Delta g$, for a target reversible inductance $L_{\hat{rev}}$ at given low-frequency peak current $\hat i_{LF}$ and a core temperature distribution $\mathcal{T}_c$. The adoption of $g_{opt}$ is optimum in the sense that a) the maximum possible manufacturing tolerance $\Delta g$ is ensured, for the resulting $N_{min}$ and b) if the actual $g \in [g_{opt}-\Delta g, g_{opt}+\Delta g]$ then the actual $L_{\hat{rev}}$ is at least the target one. This requires the initial selection of a) an ungapped core model to get its dimensions and the properties of the ferrite material and b) the winding wire gauge, from which are obtained estimations of wire temperature $T_w$ and $\mathcal{T}_c$. The following constraints should be initially specified: a) the maximum number of winding turns allowed ($N^*_{high}$), b) the number of air gaps used ($q_g$) and c) the per-unit manufacturing precision limit for the air gap length ($Tol_g$). Additionally, an unreachable lowest limit of $N_{min}$, $N^*_{low}$ should be selected in order to ultimately find $N_{min} \in (N^*_{low}, N^*_{high}]$. Inductors are specified as $q_g=1$ for being based on a core having a single air gap in the central leg while as $q_g=2$ for having the two halves of the PM core separated by an spacer, conforming so another gap divided in each of the two external legs \cite{ReporteInternoPowerModuleCores:AVS}. 
   
Although \cite{MyPaper:AVS_MR} is suitable to design inductors subjected to a certain type of non-uniform temperature distribution in the core volume, all inductors designed in this paper are based on a ferrite PM core model 
operated at a uniform $\mathcal{T}_c=\vec{100^oC}$ 
unless otherwise noted, since the other methods to be compared with only allow the hypothesis of uniform core temperature. 
Tolerance $Tol_g$ is here set to 10\% for the sake of algorithm convergence in the case of $q_g=1$ but it is disregarded elsewhere in order 
to obtain the best possible $N_{min}$. 
In all the simulation, $\Delta \Psi_{HF}$ is related to $\Delta i_{HF}$ as
\begin{align*}
\Delta i_{HF} L_{\hat {rev}} &\approx \Delta \Psi_{HF} = \frac{1}{6000} V \! s
\end{align*}
Accordingly, the target values of $L_{\hat {rev}}$ are selected in such a way that $\frac{\Delta i_{HF}}{\hat i_{LF}} \ll 1$ to be so placed in a small-signal scenario in which $L_{\hat {\Delta}}$ is very close to $L_{\hat {rev}}$ \cite{ReporteInternoDefinitionsBackground:AVS}. This is required for a further experimental assessment of $L_{\hat {rev}}$ by measuring $L_{\hat {\Delta}}$.  

\section{Comparison: Linearization of the magnetization curve and use of the Fringing factor} \label{sec:Linear_Fringing}
In this section, the design method presented in \cite{MyPaper:AVS_MR} is compared with the widely used method explained in \cite{InductorsTransformersPowerElectronics:Bossche}. 
It essentially considers $L_i=L_a=L_{\hat {\Delta}}=L_{\hat {rev}}$ as long as in any part of the core, the absolute peak induction $\hat B_{LF}+\frac{\Delta B_{HF}}{2} \leq B_{max}$. $B_{max}\approx 0.35T$ is commonly set regardless the kind of ferrite material while a core temperature $T_c=100^oC$ is also usually assumed uniform in all parts of the core. 

The number of turns $N$ is simply given by
\begin{align*}
N&=ceil \left( \frac{L_{\hat {rev}} \hat i_{LF}}{B^*_{max}A_{min}} \right) \\
B^*_{max} &=\frac{B_{max}}{1+ \frac{1}{2}\frac{\Delta i_{HF}}{\hat i_{LF}}}
\end{align*}
where $A_{min}$ is the minimum core cross-sectional area. In PM cores, $A_{c1}$ stands for the cross-sectional area of the central leg \cite{ReporteInternoPowerModuleCores:AVS} which coincides with $A_{min}$ \cite{PM-cores_made_of_magnetic_oxides_and_associated_parts:IEC61247}. As was noted before, in all simulations it holds 
$\frac{\Delta i_{HF}}{\hat i_{LF}} \ll 1$ 
and so $B^*_{max} \approx B_{max}$. 

Being the PM a three-legged core, to obtain the gap length $g$ when $q_g=1$,  
\cite{InductorsTransformersPowerElectronics:Bossche} avoids solving the nonlinear implicit equations \cite{TransformerInductorDesign:McLyman}
\begin{align}
\label{eq:Lrev_F_g}
g&=\frac{\mu_0 A_e N^2}{L_{\hat{rev}}}F-\frac{l_e}{\mu _i} \\
\label{eq:Lrev_F_F}
 F&=1+\frac{g}{\sqrt{A_{c1}}} \ln \left( \frac{2h_1}{g} \right) 
\end{align} 
by using the approach explained next. 
$F$ is the so-called fringing factor; $\mu_0$ is the vacuum permeability; $A_e$ and $l_e$ are the effective core area and core length respectively. $h_1$ is the height of the core central leg, defined for PM cores in 
\cite{ReporteInternoPowerModuleCores:AVS}. In Equation \eqref{eq:Lrev_F_g}, the approximation $F \approx (F^*-1)F^*+1$ is then used, where $F^*$ comes from Equation \eqref{eq:Lrev_F_F} when $g=g^*$. The ideal air gap $g^*$ is obtained from Equation \eqref{eq:Lrev_F_g} by making $F=1$. Note that \cite{InductorsTransformersPowerElectronics:Bossche} does not have provisions for the case $q_g=2$, 
where there are two air gaps with equal $g$ but different cross-sectional areas: the central leg area $A_{c1}$ and the external legs combined area $A_{c5}$ \cite{ReporteInternoPowerModuleCores:AVS}. To obtain better design results in those cases, this paper proposes the use of
\begin{align}
\label{eq:Lrev_F_Fmod}
 F&=\frac{1}{\frac{1}{F_1}+\frac{1}{F_5}} 
 \end{align} 
as a natural extension to the fringing factor $F$ of Equation \eqref{eq:Lrev_F_F}. The fringing factors of each air gap, $F_1$ and $F_5$ are identical to Equation \eqref{eq:Lrev_F_F} except for $F_5$ in which $A_{c1}$ is replaced by $A_{c5}$.

\begin{figure}
				\includegraphics[width=8.5cm]{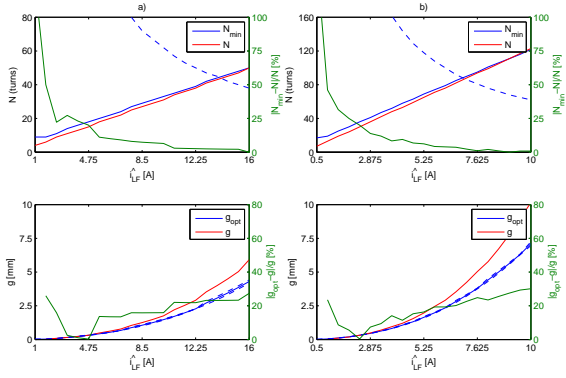} 
			\caption{$\{N_{min}, g_{opt}\}$ and $\{N, g\}$ as a function of $\hat i_{LF}$, for a constant $L_{\hat{rev}}$: a) 0.5mH, b) 2mH. N87 ferrite material, $q_g=1$.}
       \label{fig:Experiment_NvsiLF_L_1G_N87}
\end{figure}
\begin{figure}
				\includegraphics[width=8.5cm]{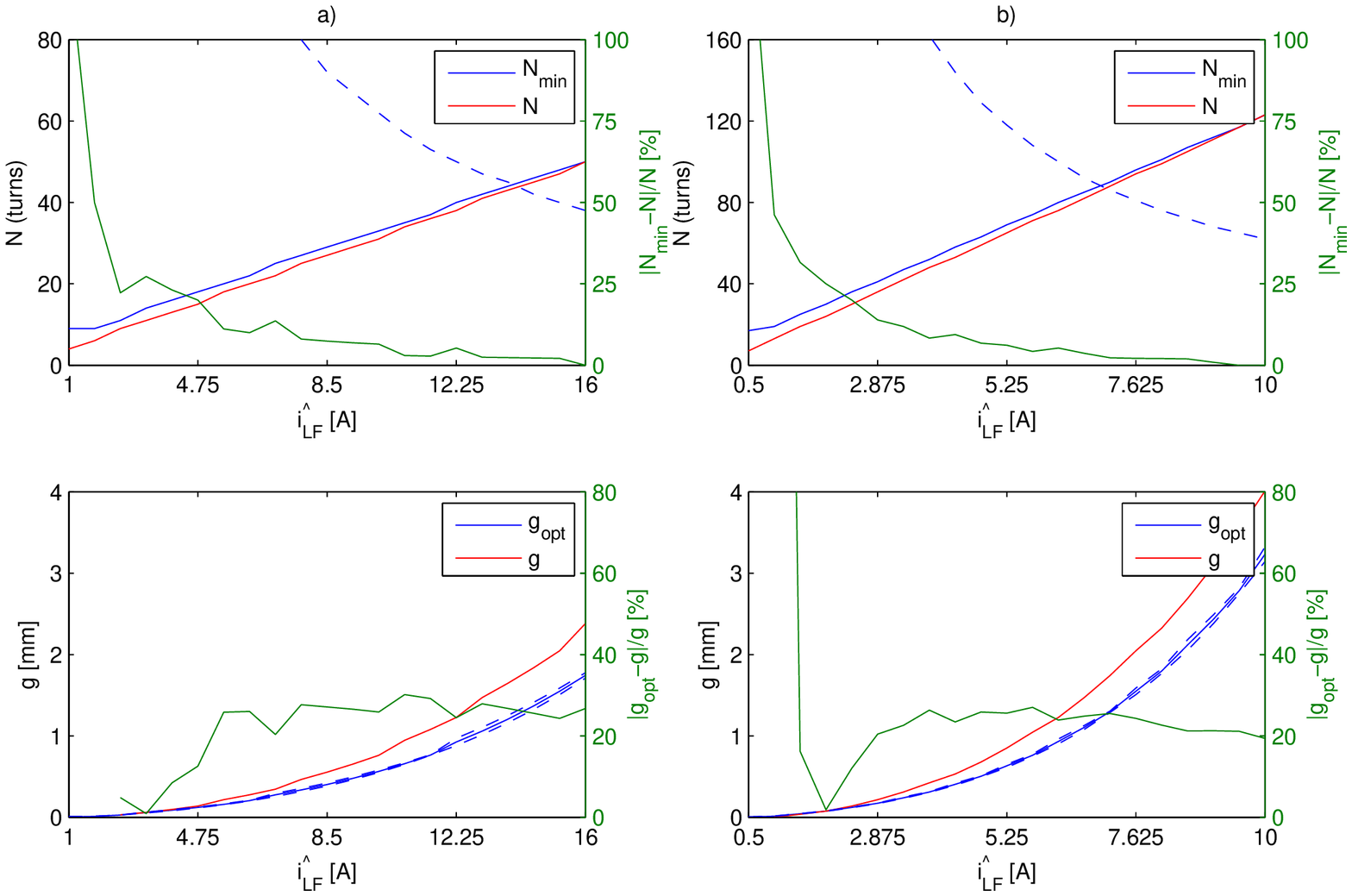} 
			\caption{$\{N_{min}, g_{opt}\}$ and $\{N, g\}$ as a function of $\hat i_{LF}$, for a constant $L_{\hat{rev}}$: a) 0.5mH, b) 2mH. N87 ferrite material, $q_g=2$.}
       \label{fig:Experiment_NvsiLF_L_N87}
\end{figure}
\begin{figure}
				\includegraphics[width=8.5cm]{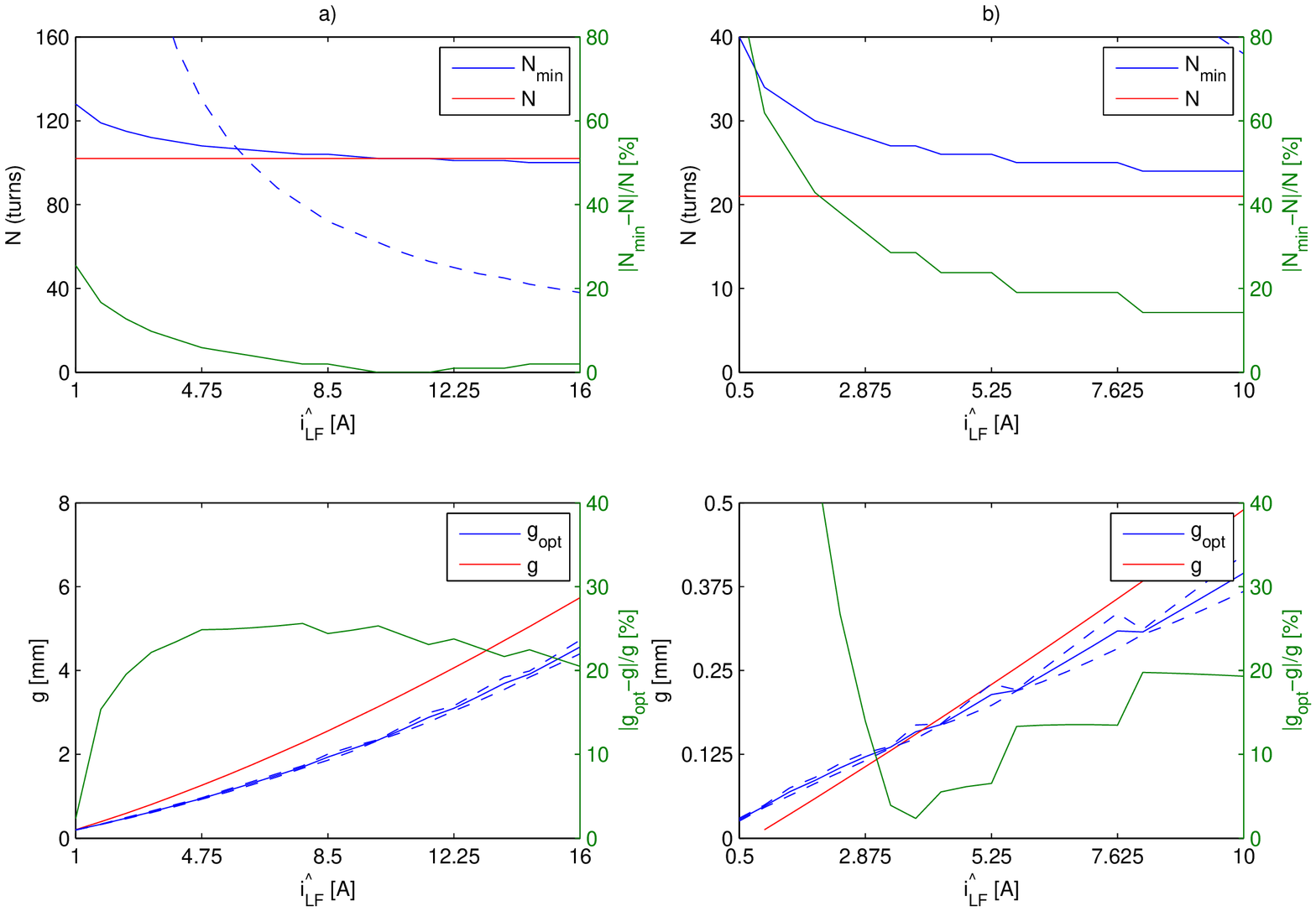} 
			\caption{$\{N_{min}, g_{opt}\}$ and $\{N, g\}$ as a function of $\hat i_{LF}$, for a constant $\frac{\Delta i_{HF}}{\hat i_{LF}}$: a) $1\%$, b) $5\%$. N27 ferrite material, $q_g=2$.}
       \label{fig:Experiment_NvsiLF_diHFiLF_N27}
\end{figure}
\begin{figure}
				\includegraphics[width=8.5cm]{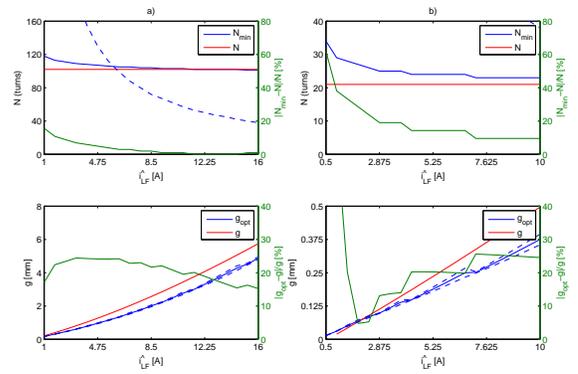} 
			\caption{$\{N_{min}, g_{opt}\}$ and $\{N, g\}$ as a function of $\hat i_{LF}$, for a constant $\frac{\Delta i_{HF}}{\hat i_{LF}}$: a) $1\%$, b) $5\%$. N87 ferrite material, $q_g=2$.}
       \label{fig:Experiment_NvsiLF_diHFiLF_N87}
\end{figure}

\begin{figure}
				\includegraphics[width=8.5cm]{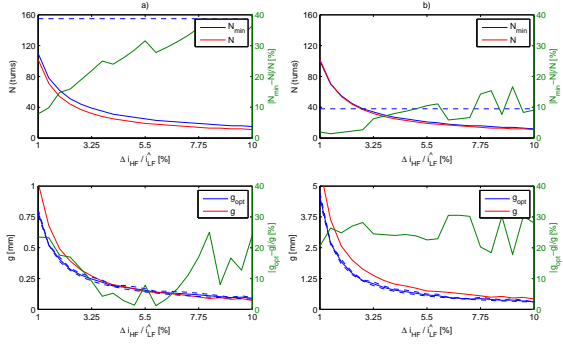} 
			\caption{$\{N_{min}, g_{opt}\}$ and $\{N, g\}$ as a function of $\frac{\Delta i_{HF}}{\hat i_{LF}}$, for a constant $\hat i_{LF}$: a) $4A$, b) $16A$. N27 ferrite material, $q_g=2$.}
       \label{fig:Experiment_NvsL_iLF_N27}
\end{figure}

\begin{figure}
       %\centering
				\includegraphics[width=8.5cm]{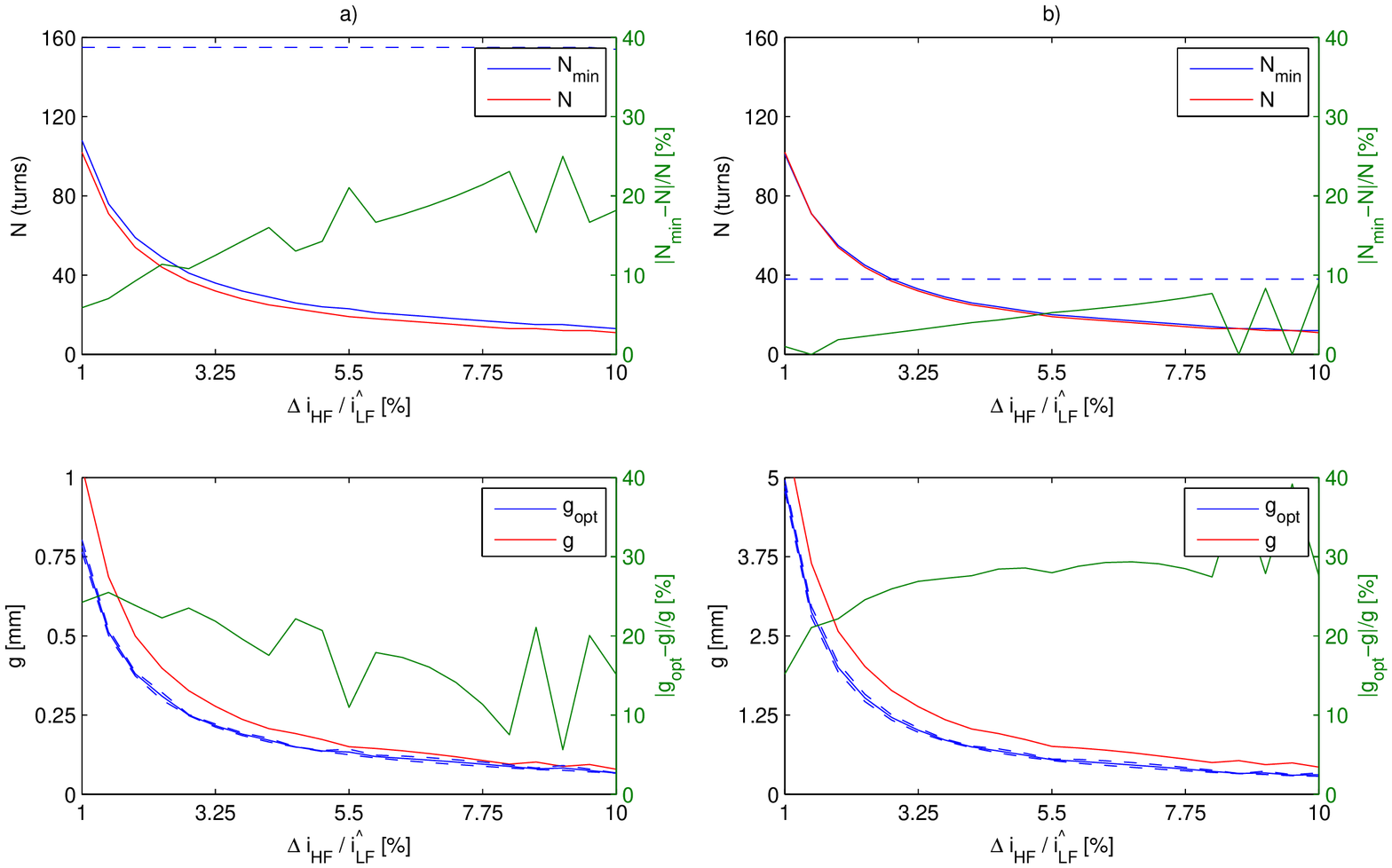} 
			\caption{$\{N_{min}, g_{opt}\}$ and $\{N, g\}$ as a function of $\frac{\Delta i_{HF}}{\hat i_{LF}}$, for a constant $\hat i_{LF}$: a) $4A$, b) $16A$. N87 ferrite material, $q_g=2$.}
       \label{fig:Experiment_NvsL_iLF_N87}
\end{figure}

In the following simulation, the resulting solutions $\{N_{min}, g_{opt}\}$ using the design algorithm of \cite{MyPaper:AVS_MR}
are compared against $\{N, g\}$ of \cite{InductorsTransformersPowerElectronics:Bossche}, for the same design specifications.
A first set of comparisons (Figures \ref{fig:Experiment_NvsiLF_L_1G_N27}-\ref{fig:Experiment_NvsiLF_L_N87}) are made as a function of the target $\hat i_{LF}$ for a constant target $L_{\hat {rev}}$: 0.5mH in Subfigures a); 2mH in Subfigures b). A second set of comparisons (Figures \ref{fig:Experiment_NvsiLF_diHFiLF_N27}-\ref{fig:Experiment_NvsiLF_diHFiLF_N87}) are made as a function of the target $\hat i_{LF}$ for a constant target $\frac{\Delta i_{HF}}{\hat i_{LF}}$: 1\% in Subfigures a); 5\% in Subfigures b). A third set of comparisons (Figures \ref{fig:Experiment_NvsL_iLF_N27}-\ref{fig:Experiment_NvsL_iLF_N87}) are made as a function of the target $\frac{\Delta i_{HF}}{\hat i_{LF}}$ for a constant target $\hat i_{LF}$: 4A in Subfigures a); 16A in Subfigures b). In all the upper parts of Subfigures a) and b), the dashed blue curves are references to the maximum number of turns $N_{max}$ that could be allocated inside the coil former for $k_u J=1.5A/mm^2$, where $k_u$ is the winding utilization factor and $J$ is the current density allowed in the wires. In all the lower parts of Subfigures a) and b), the dashed blue curves are $g_{opt} \pm \Delta g$.

The results show that $N_{min}$ tend to be higher than $N$ when $k_uJ$ is set relatively low, for example to purposely obtain low stray capacitance and/or low loss inductors. In these cases, $N_{min}$ should be a more accurate solution than $N$ because \cite{InductorsTransformersPowerElectronics:Bossche} does not consider $\mathcal{R}_{c_a}$ and $\mathcal{R}_{c_{\hat{rev}}}$ to be dissimilar and to be comparable to $\mathcal{R}_{gg}$, for determining the magnetic flux. When the number of turns is even lower, the impact of $\mathcal{R}_{go}$ can also further increase $N_{min}$ with respect to $N$. For the same reasons, the difference between $N_{min}$ and $N$ tends to disappear when $k_uJ$ increases since $\mathcal{R}_{gg}$ dominates then. Moreover, $N_{min}$ tends to be lower than $N$ for relatively large $k_uJ$ limits which enhances efficiency. $g_{opt}$ is clearly smaller than $g$ in most cases.

If the ferrite material N27 is replaced by the higher quality N87 one, $N_{min}$ tends to get closer to $N$ since $\mathcal{R}_{c_{\hat{rev}}}$ is lower and varies less from $\mathcal{R}_{c_i}$ as the target $\hat i_{LF}$ increases. For the same design specifications, the use of an N87 material leads to a lower $N_{min}$ than with an N27 material, which reduces the winding loss. However, \cite{InductorsTransformersPowerElectronics:Bossche} yields the same $N$ regardless the ferrite material employed.

\begin{figure}
       \centering
		\includegraphics[width=8.5cm]{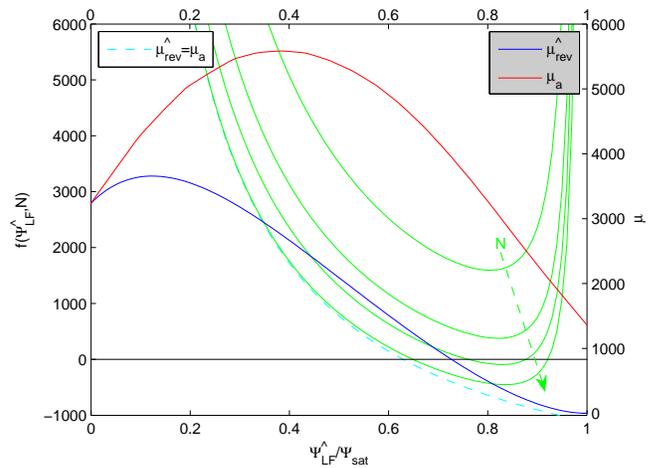}
       \caption{$\mu_a$, $\mu_{\hat{rev}}$ in $A_{min}$ and $f(\hat \Psi_{LF},N)$ as a function of $\hat \Psi_{LF}$ normalized against its saturation value $\Psi_{s}$ \cite{MyPaper:AVS_MR}}
       \label{fig:ua_urev_ffhi}
\end{figure}

It can be noted that when $N \approx N_{min}$, the trend is $g > g_{opt}$. To explain this, consider the arithmetic difference between the achievable $\frac{1}{L_{\hat{rev}}}$ for a target $\hat i_{LF}$ and the target $\frac{1}{L_{\hat{rev}}}$, as a function of $\hat \Psi_{LF}$ for a given $N$. That is referred to as $f(\hat \Psi_{LF},N)$ \cite{MyPaper:AVS_MR} and it is plotted in green lines in Figure \ref{fig:ua_urev_ffhi} for increasing values of $N$. Wherever $f(\bullet)>0$, the current $L_{\hat{rev}}$ is lower than the target one, and vice versa. Since \cite{InductorsTransformersPowerElectronics:Bossche} considers $\mu_{\hat{rev}}=\mu_a$ its corresponding $f(\bullet)$ follows the ideal dashed cyan line of Figure \ref{fig:ua_urev_ffhi}. As it usually crosses the $x$-axis at a lower $\hat \Psi_{LF}$ than in the case of the corresponding $f(\bullet)$ of \cite{MyPaper:AVS_MR}, the gap required in the former case would be larger to maintain that lower flux. It is worth mentioning that \cite{MyPaper:AVS_MR} adjusts $g_{opt}$ to place $\hat \Psi_{LF}$ in between the two points where $F(\bullet)$ crosses the $x$-axis so a bounded manufacturing tolerance in the actual $g$ can still maintain the actual $L_{\hat{rev}}$ at least equal to the target one. On the contrary, \cite{InductorsTransformersPowerElectronics:Bossche} would place $\hat \Psi_{LF}$ just in the limit and thus any minimum increase of the gap length would immediately decrease the actual $L_{\hat{rev}}$ from the target value. Likewise, \cite{InductorsTransformersPowerElectronics:Bossche} overestimates $\mu_{\hat{rev}}$ where $g \approx g_{opt}$ since there $N < N_{min}$. This is evident from Figure \ref{fig:ua_urev_ffhi} because \cite{InductorsTransformersPowerElectronics:Bossche} assumes $\mu_{\hat{rev}}$ to be always equal to $\mu_i$ but in reality, $\mu_{\hat{rev}}<\mu_i$ at the point where $\hat \Psi_{LF}$ currently operates.

As a result of those issues and according to \cite{MyPaper:AVS_MR}
\begin{align}
\label{eq:Lrev}
L_{\hat{rev}} &=\frac{N^2}{\mathcal {R}_{c_{\hat{rev}}} +\mathcal {R}_g(g)}
\end{align}
$\{N, g\}$ would yield an actual $L_{\hat{rev}}$ lower than the correspondingly predicted by $\{N_{min}, g_{opt}\}$ wherever are concurrently $g > g_{opt}$ and $N < N_{min}$.
\section{Comparison: Initial inductance and roll-off at peak current} \label{sec:Roll-off}

In this section, the design method described in \cite{MyPaper:AVS_MR} is compared with the one proposed in \cite{DC-BiasSpecificationsGappedFerriteCores:Esguerra}. It requires to initially set a target $L_i$ 
and adopt an arbitrary roll-off, $RO$, to get $L_{\hat{rev}}$ at the targeted $\hat i_{LF}$.

Once the ungapped core model is selected 
  \begin{align}
\label{eq:ue_N^2}
\mu_e &=\frac{L_{\hat {rev}}}{(1-RO)} \frac{N^2 l_e}{\mu_0 A_e} \\
RO &= \frac{L_i-L_{\hat {rev}}}{L_i} \notag 
\end{align}
where $\mu_e$ is the effective permeability of the derived gapped core to be finally employed. $N$ and $\mu_e$ are obtained simultaneously solving 
  \begin{align}
	\label{eq:Solve_N}
0 &= \frac{1-RO}{L_{\hat {rev}}} N^2 - \frac{\hat i_{LF}}{A_{min} \hat B_{LF}(\mu_{\hat{rev}})} N - \frac{l_e}{\mu_0\mu_i A_e}  \\
\mu_{\hat{rev}}&=\frac{1}{\frac{RO}{1-RO}\frac{1}{\mu_e}+\frac{1}{\mu_i}}  \notag
\end{align}
along with Equation \eqref{eq:ue_N^2}. 
The explicit function for $\mu_{\hat{rev}}\left( \hat B_{LF} \right)$ defined in \cite{ReporteInternoDefinitionsBackground:AVS}, which is also used by the inductance model of \cite{MyPaper:AVS_MR}, has to be numerically inverted to finally obtain $\hat B_{LF}\left( \mu_{\hat{rev}} \right)$ in Equation \eqref{eq:Solve_N}. 
Note that if the resulting $\mu_e$ needs to be adjusted to the nearest commercial off-the-shelf value available, then $N$ is recalculated, which may end altering $L_{\hat{rev}}$ or $\hat i_{LF}$. Otherwise, a customized air gap of length  
  \begin{align*}
 g= \left( \frac{1}{\mu_e} -  \frac{1}{\mu_i} \right) l_e 
\end{align*}
is here proposed, which is only valid for relatively small air gaps and $q_g=1$. Customized gapped core with $q_g=2$, cannot be directly handled by this design approach.

\begin{figure}
       				\includegraphics[width=8.5cm]{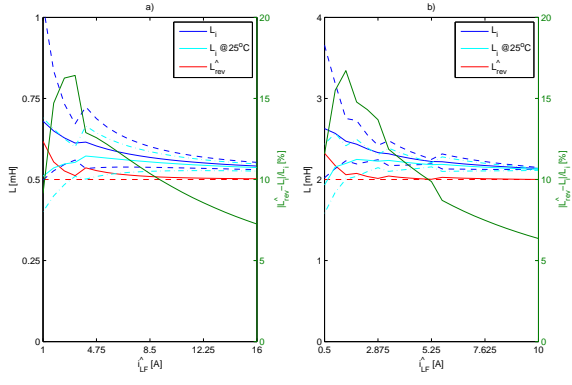} 
      			\caption{$L_i$, $L_{\hat{rev}}$ and $RO$ as a function of $\hat i_{LF}$, for a constant $L_{\hat{rev}}$: a) 0.5mH, b) 2mH. N27 ferrite material, $q_g=2$.}
       \label{fig:Experiment_NvsiLF_L_N27_2}
\end{figure}

\begin{figure}
      				\includegraphics[width=8.5cm]{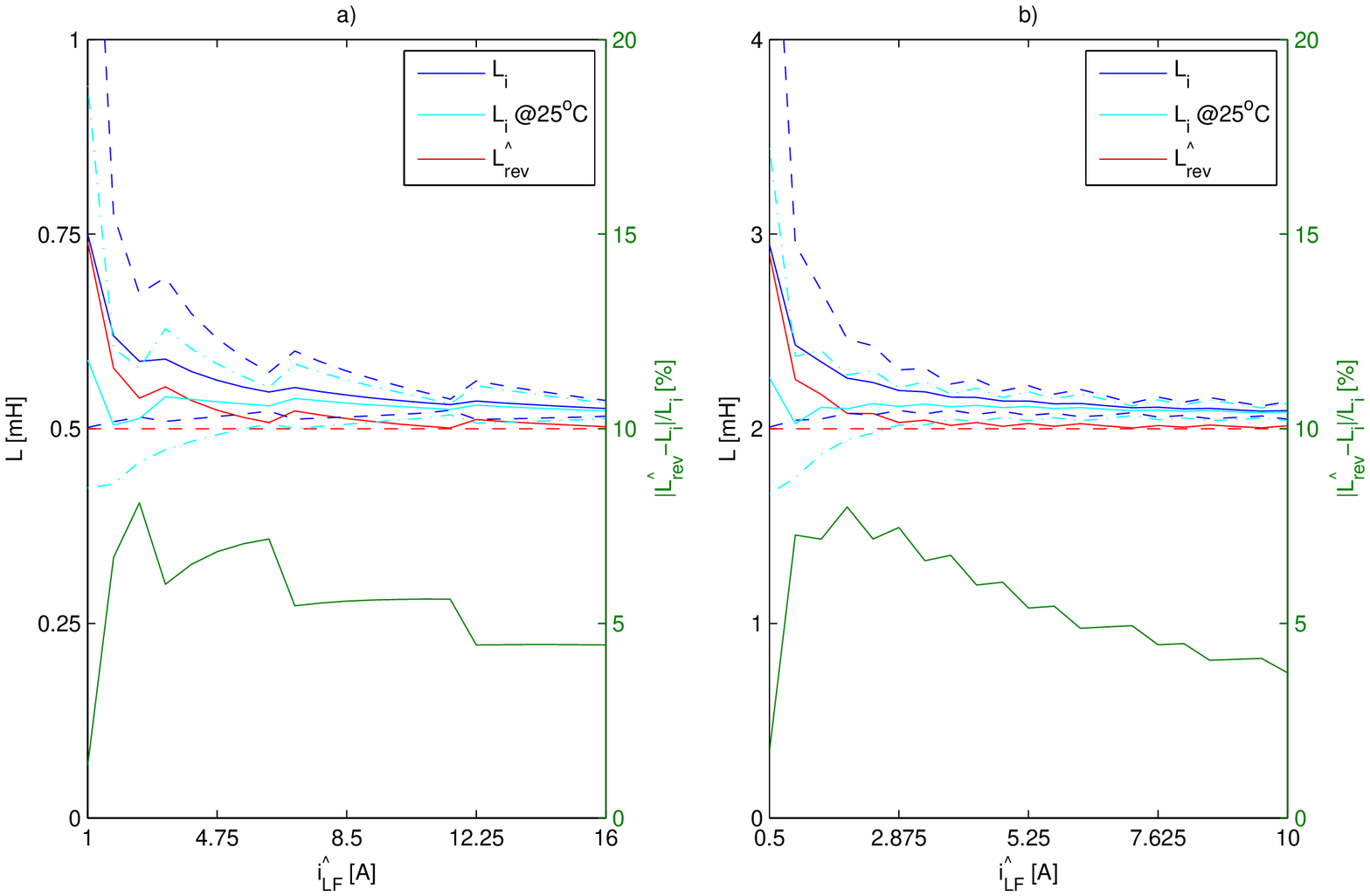} 
      			\caption{$L_i$, $L_{\hat{rev}}$ and $RO$ as a function of $\hat i_{LF}$, for a constant $L_{\hat{rev}}$: a) 0.5mH, b) 2mH. N87 ferrite material, $q_g=2$.}
       \label{fig:Experiment_NvsiLF_L_N87_2}
\end{figure}

\begin{figure}
      				\includegraphics[width=8.5cm]{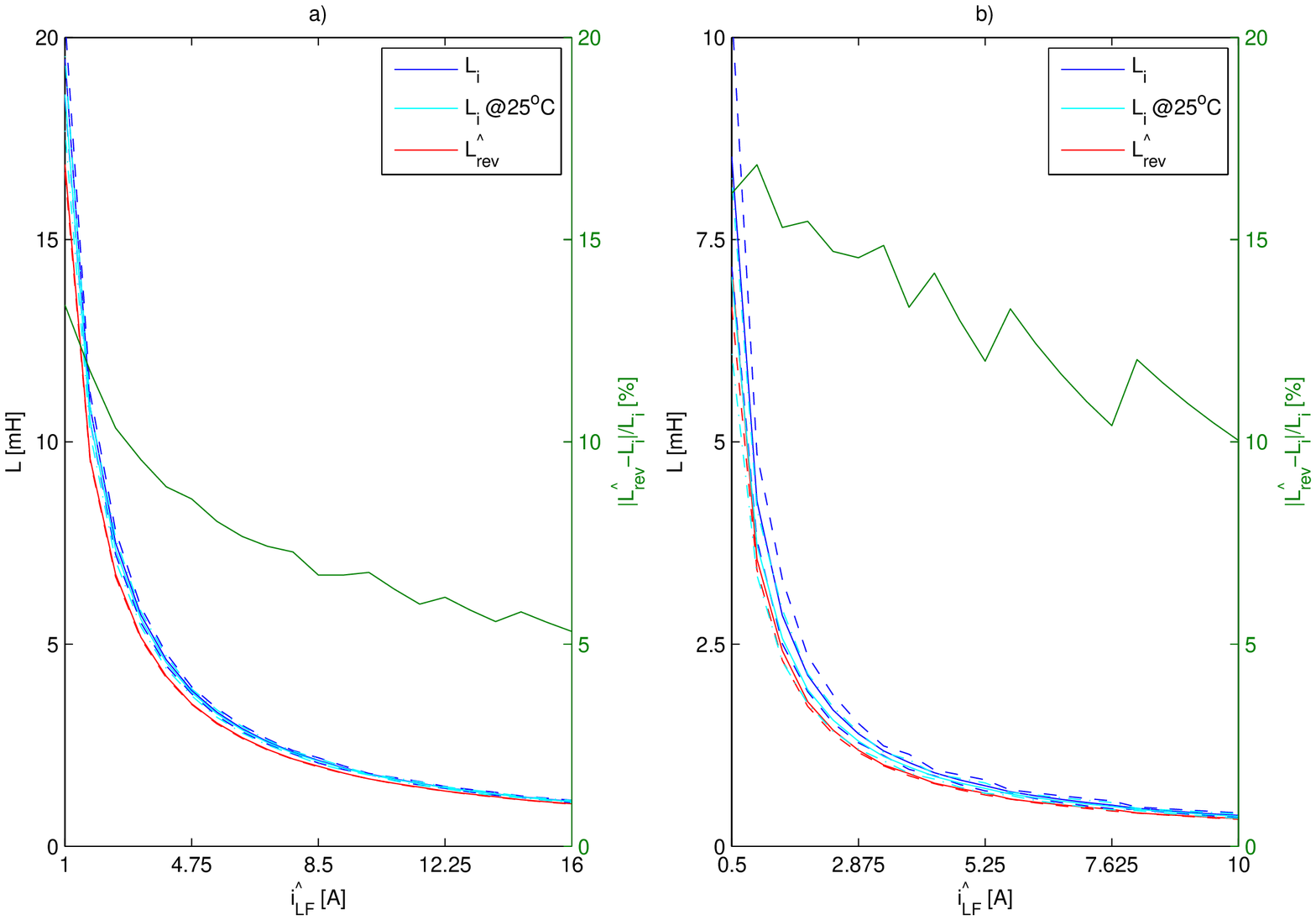} 
      			\caption{$L_i$, $L_{\hat{rev}}$ and $RO$ as a function of $\hat i_{LF}$, for a constant $\frac{\Delta i_{HF}}{\hat i_{LF}}$: a) $1\%$, b) $5\%$. N27 ferrite material, $q_g=2$.}
       \label{fig:Experiment_NvsiLF_diHFiLF_N27_2}
\end{figure}

\begin{figure}
      				\includegraphics[width=8.5cm]{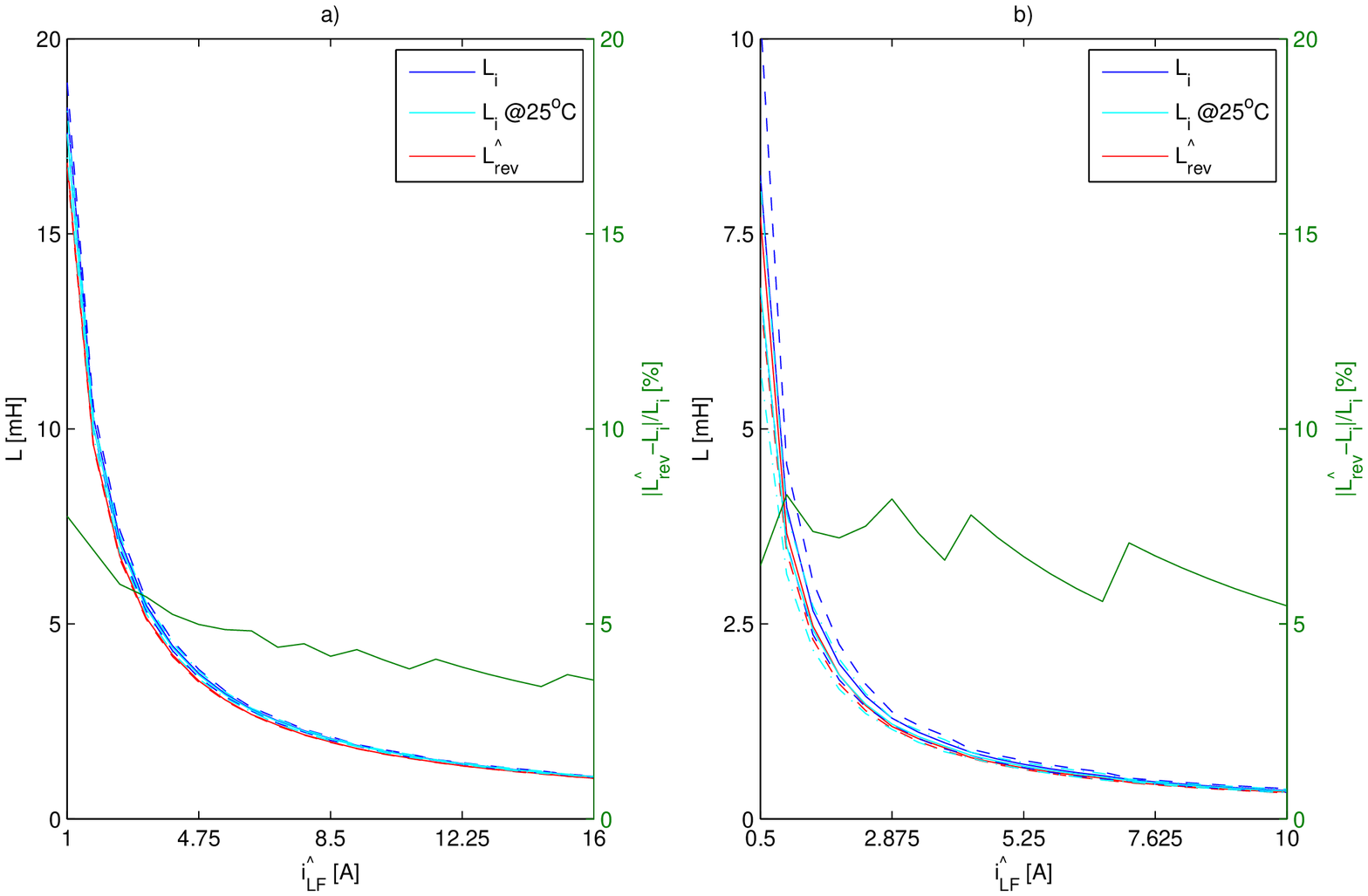} 
      			\caption{$L_i$, $L_{\hat{rev}}$ and $RO$ as a function of $\hat i_{LF}$, for a constant $\frac{\Delta i_{HF}}{\hat i_{LF}}$: a) $1\%$, b) $5\%$. N87 ferrite material, $q_g=2$.}
       \label{fig:Experiment_NvsiLF_diHFiLF_N87_2}
\end{figure}

\begin{figure}
       				\includegraphics[width=8.5cm]{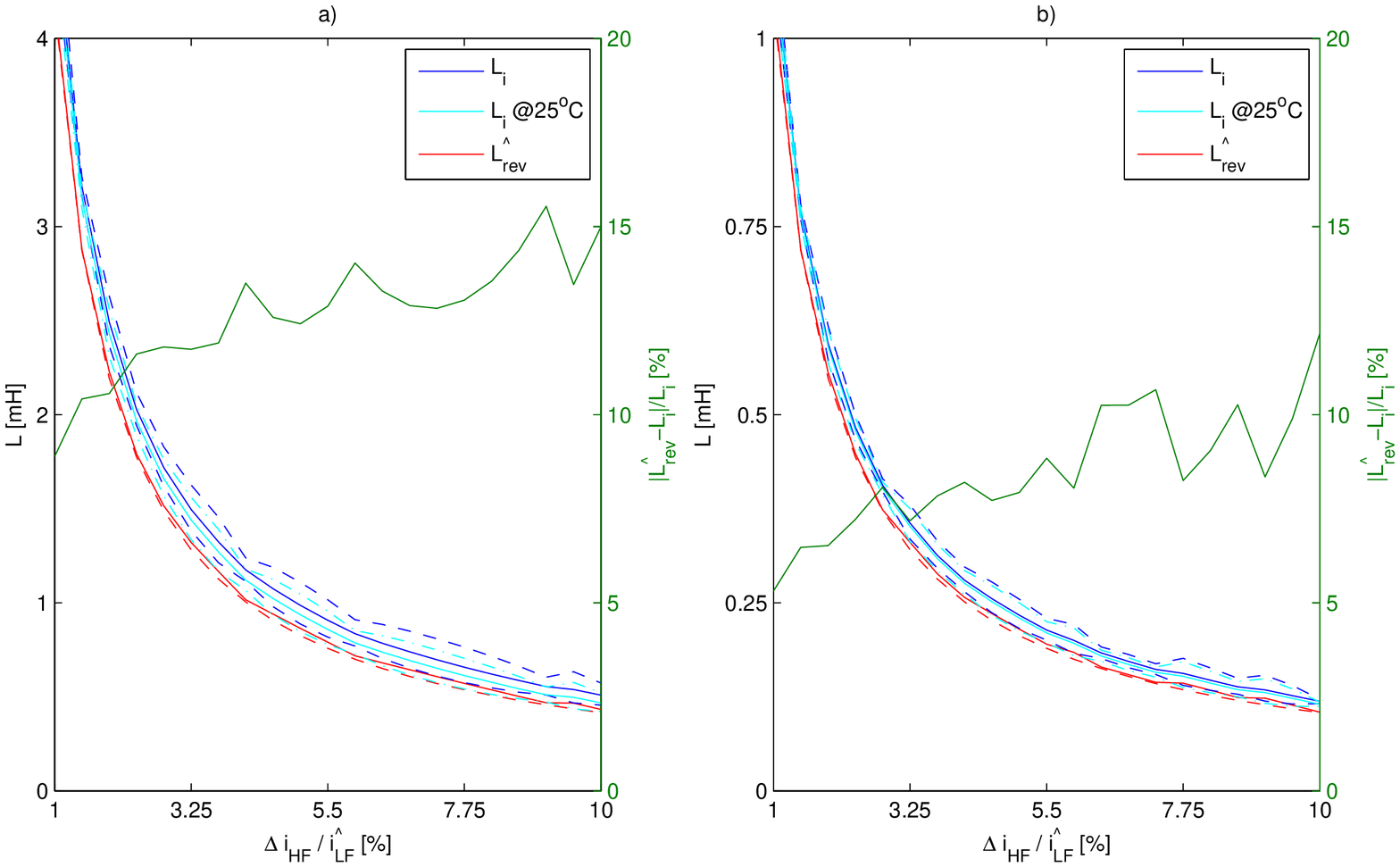} 
      			\caption{$L_i$, $L_{\hat{rev}}$ and $RO$ as a function of $\frac{\Delta i_{HF}}{\hat i_{LF}}$, for a constant $\hat i_{LF}$: a) $4A$, b) $16A$. N27 ferrite material, $q_g=2$.}
       \label{fig:Experiment_NvsL_iLF_N27_2}
\end{figure}

\begin{figure}
      				\includegraphics[width=8.5cm]{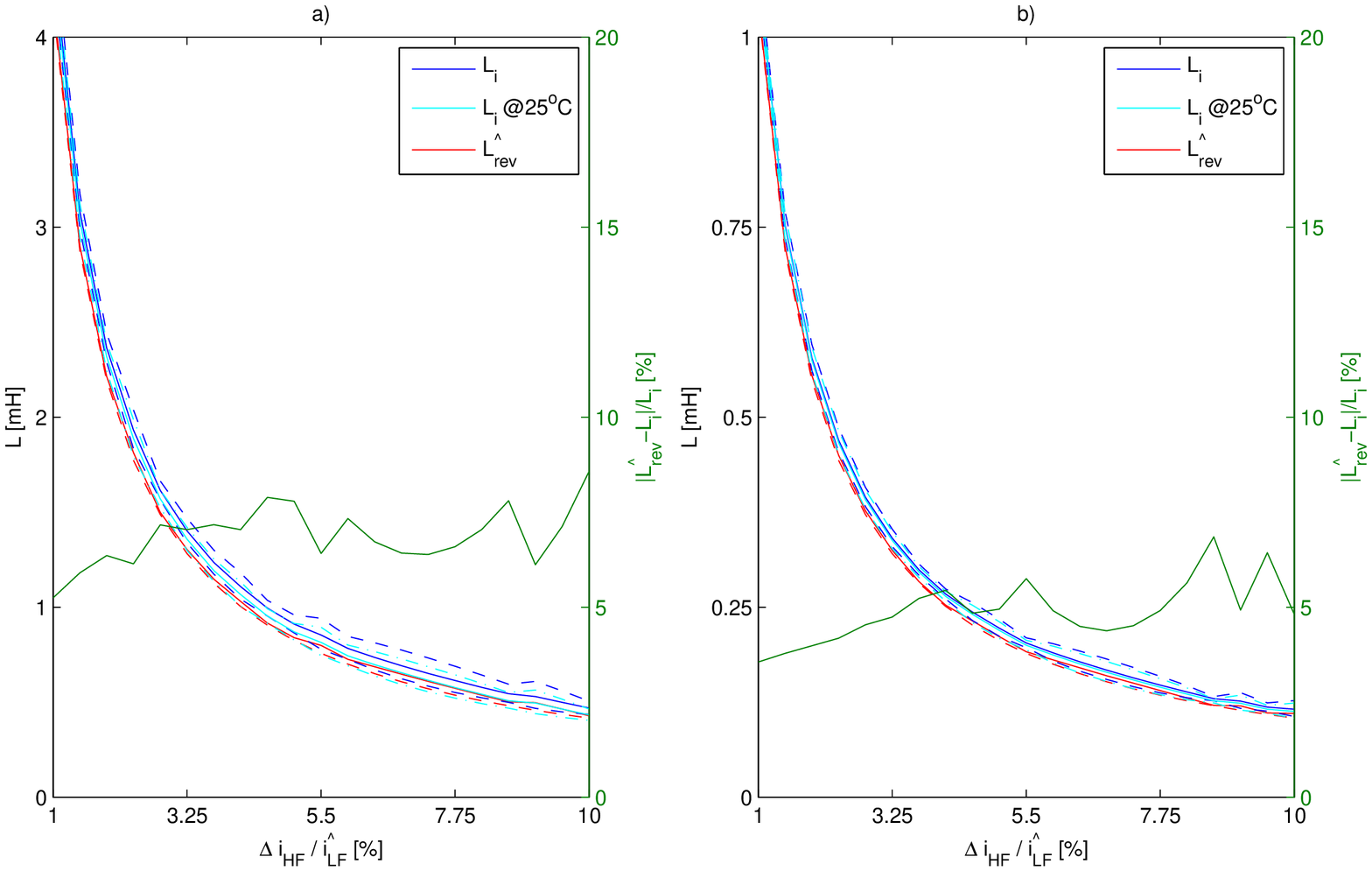} 
      			\caption{$L_i$, $L_{\hat{rev}}$ and $RO$ as a function of $\frac{\Delta i_{HF}}{\hat i_{LF}}$, for a constant $\hat i_{LF}$: a) $4A$, b) $16A$. N87 ferrite material, $q_g=2$.}
       \label{fig:Experiment_NvsL_iLF_N87_2}
\end{figure}

The main purpose of this section is not to compare absolute results of both design methods, like was done in the previous section. It is rather oriented to conceptually discuss the rationale of \cite{DC-BiasSpecificationsGappedFerriteCores:Esguerra}. To this end, $L_i$ and $L_{\hat {rev}}$ are calculated using \cite{MyPaper:AVS_MR} alike in the previous section. The resulting $RO$ are then examined. 
Figures \ref{fig:Experiment_NvsiLF_L_N27_2}-\ref{fig:Experiment_NvsiLF_diHFiLF_N87_2} show in solid lines the resulting $L_i$ and $L_{\hat{rev}}$ using nominal parameters $\{N_{min}, g_{opt}, A_L\}$ 
as a function of the target $\hat i_{LF}$, where $A_L$ is the nominal inductance factor of the ungapped core. In Figures \ref{fig:Experiment_NvsiLF_L_N27_2}-\ref{fig:Experiment_NvsiLF_L_N87_2}, constant target $L_{\hat{rev}}$: 0.5mH in subfigure a); 2mH in subfigure b) are used. In Figures \ref{fig:Experiment_NvsiLF_diHFiLF_N27_2}-\ref{fig:Experiment_NvsiLF_diHFiLF_N87_2}, constant $\frac{\Delta i_{HF}}{\hat i_{LF}}$: $1\%$ in subfigure a); $5\%$ in subfigure b) are employed. In Figures \ref{fig:Experiment_NvsL_iLF_N27_2}-\ref{fig:Experiment_NvsL_iLF_N87_2}, $L_i$ and $L_{\hat{rev}}$ are obtained as a function of $\frac{\Delta i_{HF}}{\hat i_{LF}}$ for constant $\hat i_{LF}$: 4A in subfigure a); 16A in subfigure b). For all figures, $T_{c}=25^oC $ in cyan lines. Upper and lower dashed blue and cyan lines correspond to the limit variations on $L_i$ when it is calculated with parameters $\{N_{min}, g_{opt} - \Delta g, A_{Lmax}\}$ and $\{N_{min}, g_{opt} + \Delta g, A_{Lmin}\}$, respectively. $A_{Lmax}$ and $A_{Lmin}$ are the tolerance limits of $A_L$. The red dashed lines are the limits on $L_{\hat{rev}}$ calculated under the same extreme parameters as before, which certainly coincide with the target $L_{\hat{rev}}$. 
The resulting associated $RO$ is then obtained with values of $L_i$ and $L_{\hat{rev}}$ from the solid lines in blue and red respectively.

The $RO$ tend to be smaller where ferrite material N87 rather than N27 is employed which indicates that the former one would yield inductors with more stable inductance as the existing peak current increases from zero to the rated $\hat i_{LF}$. For both materials however, inductors rated at higher $\hat i_{LF}$ or $L_{\hat{rev}}$ also tend to have an smaller $RO$ since the air gap (linear) mandates over the core reluctance (non-linear) in those situations. 
In light of these simulations it can be concluded that, if the inductor is to be designed with the minimum number of turns in mind, the resulting $RO$ is highly dependent on the target $L_{\hat{rev}}$, $\hat i_{LF}$ and the ferrite material selected. However, the actual $RO$ can only be determined once the design is over. Thus, an arbitrary adoption of $RO$ at the beginning of the design process, as the approach in \cite{DC-BiasSpecificationsGappedFerriteCores:Esguerra} demands, it would yield in most cases a sub optimal design. 

\section{Behaviour of inductors beyond specifications} \label{sec:Beyond_specifications} 

\begin{figure}
      				\includegraphics[width=8.5cm]{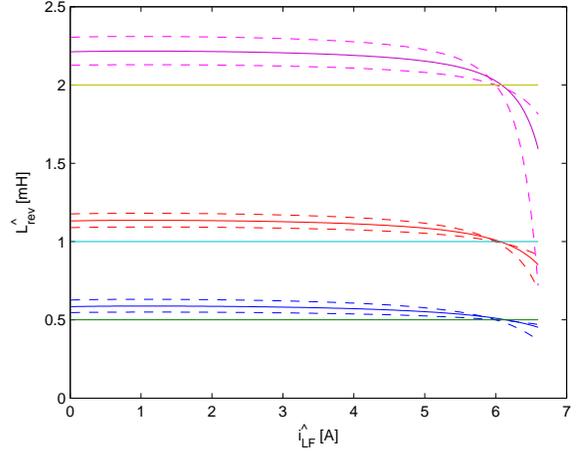} 
      			\caption{$L_{\hat{rev}}$ as a function of $\hat i_{LF}$, varying from $0A$ up to a $10\%$ more than the design target $\hat i_{LF}=6A$, for three different values of target inductance}
       \label{fig:Experiment_LrevvsiLF_robust}
\end{figure}

Figure \ref{fig:Experiment_LrevvsiLF_robust} shows the evolution of $L_{\hat{rev}}$, as the current value of $\hat i_{LF}$ through the inductor increases from 0A up to a 10\% more than the rated current $\hat i_{LF}=6A$, for a design target $L_{\hat{rev}}$ equal to $2mH$, $1mH$ and $0.5mH$. While the current $\hat i_{LF}$ is lower than the design target $\hat i_{LF}$, the upper and lower dashed lines correspond to the limit situations $\{g_{opt} - \Delta g, A_{Lmax}\}$ and $\{g_{opt} + \Delta g, A_{Lmin}\}$ respectively; after that, their relative positions invert. The solid line between them is the resulting nominal inductance with $\{g_{opt}, A_{L}\}$. As it is guaranteed, the dashed lines converge to the design target $L_{\hat{rev}}$ when $\hat i_{LF}$ matches the design target $\hat i_{LF}$ and, as is expected, the more $g$ approaches to $g_{opt}+\Delta g$, the more robust is the inductor when the design target $\hat i_{LF}$ is surpassed. This is achieved at the expense of obtaining an inductance that, though higher than the design target $L_{\hat{rev}}$, is lower than the nominal while the current $\hat i_{LF}$ is below the design target $\hat i_{LF}$.  
\begin{figure}
      				\includegraphics[width=8.5cm]{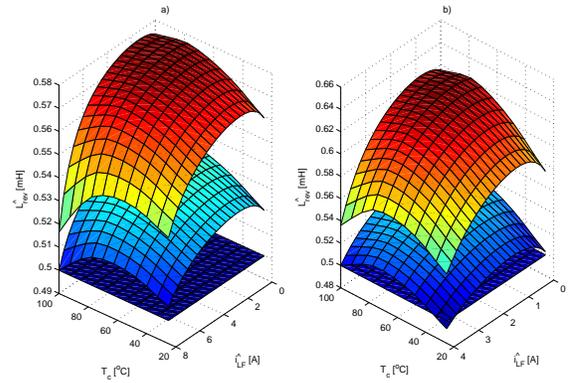} 
      			\caption{$L_{\hat{rev}}$ as a function of current $\hat i_{LF}$ and $\mathcal{T}_c$. Design $\{L_{\hat{rev}},\mathcal{T}_c\}=\{0.5mH,\vec{100^oC}\}$; design $\hat i_{LF}$ equal to a) 8A and b) 4A.}
       \label{fig:Experiment_LrevvsiLFvsTcore2}
\end{figure}
The previous analysis was done under the assumption of a uniform and constant core temperature $\mathcal{T}_c=\vec{100^oC}$. Figure \ref{fig:Experiment_LrevvsiLFvsTcore2}-a) shows the behaviour of $L_{\hat{rev}}$ when $\mathcal{T}_c$ varies from $\vec{100^oC}$ down to $\vec{25^oC}$. The plane parallel to the $x-y$ axes is the design target inductance $L_{\hat{rev}}=0.5mH$. The lower curved surface is the lower limit of inductance, which touches the plane of target inductance only at the design point $\{\hat i_{LF}=8A, \mathcal{T}_{c}=\vec{100^oC}\}$. The upper surface is the nominal inductance, always well above the other two surfaces. Although in this particular design made at $\mathcal{T}_c=\vec{100^oC}$ none of the curved surfaces trespasses the plane of target inductance while the current $\mathcal{T}_c$ decreases from that design temperature, this cannot be ensured to be so for any design. The behaviour of another inductor based on the same core model with the same $L_{\hat{rev}}$ but rated at $\hat i_{LF}=4A$ is depicted in Figure \ref{fig:Experiment_LrevvsiLFvsTcore2}-b). It reveals that the minimum value of inductance can take values below the design target as $\mathcal{T}_c$ approaches $\vec{25^oC}$. That may or may not be a problem but it suggests doing another design attempt at the lowest temperature of interest and then to test whether that inductor complies with the required inductance all through the temperature span.       

\section{Conclusion} \label{sec:Conclusion}
The comparative results obtained from the simulation on the PM core TDK-EPCOS 62/49 have shown that the inductance model and the design method developed in \cite{MyPaper:AVS_MR} has clear advantages over the other common design approaches here examined. Being grounded on a sufficiently detailed inductance model, the optimized design method brings to the designer much more confidence in obtaining an inductor with a guaranteed inductance at the design specifications of peak current and core temperature. Also it proves that the inevitable manufacturing tolerances in the air gap length, if are kept below a certain limit, are not going to degrade the specifications of the designed inductor. Although \cite{MyPaper:AVS_MR} requires a more complex modelling of the inductor and more number of steps are devoted in finding $N_{min}$ and $g_{opt}$, all core parameters are readily available from the manufacturer's datasheet while the model and the design strategy can be easily implemented under any scientific programming language, such as Matlab or Scilab. Once the inductance model is numerically implemented, it can also be conveniently employed for a given inductor, designed according to any methodology, to verify its inductance evolution under typical or limiting operational scenarios as well as predict behaviours beyond specifications. Future works are expected in assessing inductors based on other core shapes and performing experimental measurements of inductance over prototype inductors designed with \cite{MyPaper:AVS_MR}.           

\appendices

\section{References to variables and parameters} \label{sec:Appendix}
  
\begin{table}[ht]
\caption{List of Variables and Parameters} 
\centering 
\begin{tabular}{c c c}
\hline\hline 
Symbol & Definition & Ref. \\ [0.5ex] 
\hline 
$\hat B_{LF}$ & Low-frequency magnetic induction at $\hat \Psi_{LF}$ & \cite{ReporteInternoDefinitionsBackground:AVS} \\
$\Delta B_{HF}$ & High-frequency incremental magnetic induction & \cite{ReporteInternoDefinitionsBackground:AVS} \\
$\hat i_{LF}$ & Low-frequency peak current & \cite{ReporteInternoDefinitionsBackground:AVS} \cite{MyPaper:AVS_MR} \\
$\Delta i_{HF}$ & High-frequency incremental inductor current & \cite{ReporteInternoDefinitionsBackground:AVS} \cite{MyPaper:AVS_MR}\\   
$L_i$ & Initial inductance & \cite{ReporteInternoDefinitionsBackground:AVS} \cite{MyPaper:AVS_MR} \\
$L_a$ & Amplitude inductance at $\hat i_{LF}$ & \cite{ReporteInternoDefinitionsBackground:AVS} \cite{MyPaper:AVS_MR}\\
$L_{\hat {\Delta}}$ & Incremental inductance at $\hat i_{LF}$ & \cite{ReporteInternoDefinitionsBackground:AVS} \cite{MyPaper:AVS_MR}\\ 
$L_{\hat{rev}}$ & Reversible inductance at $\hat i_{LF}$ & \cite{ReporteInternoDefinitionsBackground:AVS} \cite{MyPaper:AVS_MR} \\ 
$\mu_i$ & Initial permeability & \cite{ReporteInternoDefinitionsBackground:AVS} \\
$\mu_a$ & Amplitude permeability at $\hat B_{LF}$ & \cite{ReporteInternoDefinitionsBackground:AVS} \\
$\mu_{\hat{rev}}$ & Reversible permeability at $\hat B_{LF}$ & \cite{ReporteInternoDefinitionsBackground:AVS} \\
$\mathcal{R}_{c_i}$ & Initial core reluctance & \cite{ReporteInternoDefinitionsBackground:AVS} \\
$\mathcal{R}_{c_a}$ & Amplitude core reluctance at $\hat \Psi_{LF}$ & \cite{ReporteInternoDefinitionsBackground:AVS} \\
$\mathcal{R}_{c_{\hat{rev}}}$ & Reversible core reluctance at $\hat \Psi_{LF}$ & \cite{ReporteInternoDefinitionsBackground:AVS} \\
$\mathcal{R}_{gg}$ & Air gap reluctance due to length $g$ & \cite{MyPaper:AVS_MR} \\
$\mathcal{R}_{go}$ & Residual air gap reluctance & \cite{MyPaper:AVS_MR} \\
$\hat \Psi_{LF}$ & Low-frequency magnetic flux at $\hat i_{LF}$ & \cite{ReporteInternoDefinitionsBackground:AVS} \cite{MyPaper:AVS_MR}\\
$\Delta \Psi_{HF}$ & High-frequency incremental magnetic flux & \cite{ReporteInternoDefinitionsBackground:AVS} \cite{MyPaper:AVS_MR}\\
$\mathcal{T}_c$ & Vector of core temperature distribution & \cite{MyPaper:AVS_MR}\\ [1ex] 
\hline 
\end{tabular}
\label{table:param_var} 
\end{table}
In Table \ref{table:param_var} a list of variables and parameters referred in this paper, which are employed by the inductance model and the design method of \cite{MyPaper:AVS_MR}, is presented. In each cited reference, the definition of the parameter is given and explained into an appropriate context.

\section*{Acknowledgment}

The  first author wants to thank Dr. Hernan Haimovich for his guidance and constructive suggestions.

\ifCLASSOPTIONcaptionsoff
  \newpage
\fi

\end{document}